\documentclass[aps,prep,preprintnumbers,amsmath,amssymb,nofootinbib]{revtex4}
\usepackage{amssymb}
\usepackage{amsmath}
\usepackage{bm}
\usepackage{graphicx}
\usepackage{epsfig}
\begin{document}

\title{Scalar fluctuations of the scalar metric during inflation from a non-perturbative 5D large-scale repulsive gravity model}
\author{$^{1}$ Jos\'e Edgar Madriz Aguilar \thanks{E-mail address: madriz@mdp.edu.ar}, $^{1}$ Luz M. Reyes ,
$^{1}$ Claudia Moreno \thanks{E-mail address: claudia.moreno@cucei.udg.mx}
 and $^{2,3}$ Mauricio Bellini
\thanks{E-mail address: mbellini@mdp.edu.ar}, }
\affiliation{$^{1}$ Departamento de Matem\'aticas, Centro Universitario de Ciencias Exactas e ingenier\'{i}as (CUCEI),
Universidad de Guadalajara (UdG), Av. Revoluci\'on 1500 S.R. 44430, Guadalajara, Jalisco, M\'exico.  \\
E-mail: edgar.madriz@red.cucei.udg.mx, luzmarinareyes@gmail.com, claudia.moreno@cucei.udg.mx\\
and \\
$^{2}$ Departamento de F\'isica, Facultad de Ciencias Exactas y Naturales, Universidad Nacional de Mar del Plata (UNMdP),
Funes 3350, C.P. 7600, Mar del Plata, Argentina \\
E-mail: mbellini@mdp.edu.ar\\
$^{3}$ Instituto de Investigaciones F\'{i}sicas de Mar del Plata
(IFIMAR)- Consejo Nacional de Investigaciones Cient\'{i}ficas y
T\'ecnicas (CONICET) Argentina.}

\begin{abstract}
We develop a non-perturbative formalism for scalar metric
fluctuations from a 5D extended version of general relativity in
vacuum. In this work we concentrate our efforts on calculations
valid on large cosmological scales, which are the dominant during
the inflationary phase of the universe. The resulting metric in
this limit is obtained after implementing a planar coordinate
transformation on a 5D Ricci-flat metric solution. We calculate
the spectrum of these fluctuations with an effective 4D
Schwarzschild-de Sitter spacetime on cosmological scales, which is
obtained after we make a static foliation on the non-compact extra
coordinate. Our results show how the squared metric fluctuations
of the primordial universe become scale invariant with the
inflationary expansion.
\end{abstract}

\pacs{04.20.Jb, 11.10.kk, 98.80.Cq, 04.30.-w, 04.62.+v}
\maketitle

\vskip .5cm

Keywords: scalar metric fluctuations, cosmological inflation, 5D non-compact Kaluza-Klein gravity, Schwarzschild-de-Sitter metric.

\section{Introduction}

In the last years the study of the early universe has raised the
interest of cosmologists due to some new issues on this topic as
new inflationary cosmologies, quantum gravity, etc. Currently
inflation is a well grounded theory having an experimental status,
that has a good interface with high-energy physics and provides
various approaches to topical problems of the fundamental physics,
specifically to the problem of divergence in a quantum theory or
the singularity problem in the general theory of relativity.
Furthermore, in some way it leads to the problem of
quantum-gravitational effects and their adequate description.
These reasons, among others, make the investigation of the early
Universe of particular importance. Inflationary cosmology is one
of the paradigms that will be confirmed with new observational
data coming from the
 PLANCK satellite. Inflation has become the standard paradigm for
explaining the homogeneity and the isotropy of our observed
Universe \cite{1,2}. During this epoch the energy density of the
Universe was dominated by some scalar field (the inflaton), with
negligible kinetic energy density, in such a way that its
corresponding vacuum energy density is responsible for the
exponential growth of the scale factor of the universe. Along this
phase a small and smooth region of the order of size of the Hubble
radius, grew so large that it easily encompassed the comoving
volume of the entire presently observed Universe, and consequently
the observable universe become so homogeneous and isotropic,
spatially. Moreover, it is now clear that the structure in the
Universe has its origin primarily with an almost scale-invariant
super-horizon curvature perturbation. Unfortunately, there are
plenty of inflationary models the majority of them in good
concordance with observations, but none free of problems, as for
example the transplanckian problem, the hierarchy problem, etc
\cite{bran}. This has lead cosmologists to look for some new
theoretical alternatives \cite{nrr}.
In particular the resort to extra dimensions has been of some success.\\

The extension of the 4D spacetime to $N(\geq 5)D$ manifolds is the
preferred route to a unification of the interactions of particle
physics with gravity. The basic extension to 5D describes a spin-2
graviton, a spin-1 photon and a spin-0 scalaron, which can be
related, respectively, to the Einstein, Maxwell and Higgs fields
\cite{we}. In the cosmological context the extra dimension is
already known to be of great importance. There is a class of 5D
cosmological models which reduce to the usual four dimensional
ones, on hypersurfaces defined by establishing a foliation on the
fifth extra coordinate. In these models the matter is
explained as a consequence of the geometry in five dimensions.\\

The idea that the equation of state for matter is $\omega_m=-1$
(and not $\omega_m=0$) was previously explored in some works using
some ideas of the Space-Time-Matter (STM) theory, from the
gravitational \cite{..} and cosmological \cite{...} points of
view. The origin of gravitational waves has been also considered
for a large-scale repulsive gravitational field from a 5D vacuum
\cite{....}. In this work we develop a non-perturbative formalism
for scalar metric fluctuations from a 5D extended Schwarzschild-de
Sitter (SdS) cosmological metric. It can be seen that on
cosmological scales gravity becomes repulsive and these
fluctuations behave as a longitudinal gauge \cite{vi}. Finally, we
study the spectrum for the squared fluctuations of the metric.

\section{The 5D field equations}

On the coordinate chart $\lbrace
T,R,\theta,\phi,\psi\rbrace$, let us consider the 5D line
element
\begin{equation}\label{a1}
dS_{5}^{2}=\left(\frac{\psi}{\psi_0}\right)^{2}\left[c^{2}f(R)dT^2-\frac{dR^2}{f(R)}-R^2(d\theta^2+sin^2\theta
d\phi^2)\right]-d\psi^2,
\end{equation}
describing a Ricci-flat metric \cite{ga1}. Here,
$f(R)=1-[(2G\zeta\psi_0/(Rc^2)]-(R/\psi_0)^2$ is a dimensionless
metric function, $\psi$ is the space-like and non-compact fifth
extra coordinate\footnote{In our notation conventions henceforth,
latin indices $a,b=$ run from 0 to 4, whereas the rest of latin
indices $i,j,n,l,...=$ run from 1 to 3.}. This metric is an
extension to 5D spaces of the 4D SdS metric. $T$ is a time-like
coordinate, $c$ is denoting the speed of light, $R,\theta,\phi$
are the usual spherical polar coordinates, $\psi_0$ is an
arbitrary constant with length units and the constant parameter
$\zeta$ has units of $(mass)(length)^{-1}$. The metric (\ref{a1})
is static, however, it can be written on a dynamical coordinate
chart $\lbrace t,r,\theta,\phi,\psi\rbrace$ by implementing the
planar coordinate transformation \cite{pc}
\begin{equation}\label{a2}
R=ar\left[1+\frac{G\zeta\psi_0}{2ar}\right]^2,\quad T=t+H\int^r dR\frac{R}{f(R)}\left(1-\frac{2G\zeta\psi_0}{R}\right)^{-1/2},
\end{equation}
$a(t)=a_0e^{t/\psi_0}$ being the scale factor. After doing so, the
line element (\ref{a1}) may be expressed in terms of the conformal
time $\tau$ in the form
\begin{equation}\label{a3}
  dS_{5}^{2} = \left(\frac{\psi}{\psi_0}\right)^2 \left[F(\tau,r)d\tau^2 - J(\tau,r)(dr^2+r^2(d\theta^2+sin^2\theta
  d\phi^2))\right]-d\psi^2,
\end{equation}
where the metric functions $F(\tau,r)$ and $J(\tau,r)$ are given by
\begin{equation}\label{a4}
F(\tau,r)=a^2(\tau)\left[1-\frac{G\zeta\psi_0}{2a(\tau)r}\right]^2\left[1+\frac{G\zeta\psi_0}{2a(\tau)r}\right]^{-2},\quad
J(\tau,r)=a^2(\tau)\left[1+\frac{G\zeta\psi_0}{2a(\tau)r}\right]^4,
\end{equation}
with $d\tau=a^{-1}(\tau)dt$ and $a(\tau)=-\psi_0/\tau$, so that
the constant Hubble parameter satisfies
\begin{equation}\label{a5}
H=\frac{1}{\psi_0}=a^{-2}\frac{da}{d\tau} .
\end{equation}
As was shown in \cite{ga1}, for certain values of $\zeta$ and
$\psi_0$, the metric in (\ref{a1}) has two natural horizons. The
inner one is the analogous to the Schwarzschild horizon and the
external one is the analogous to the Hubble horizon. In the metric
in (\ref{a3}) these horizons may be written in the new
dynamical coordinate chart.\\

The non-perturbative metric fluctuations of the background (\ref{a3}), are introduced through the perturbed line element
\begin{equation}\label{a6}
\left.dS_{5}^{2}\right|_{pert}=\left(\frac{\psi}{\psi_0}\right)^2 \left[F(\tau,r)e^{2\Phi}d\tau^2 - J(\tau,r)e^{-2\Phi}(dr^2+r^2(d\theta^2+sin^2\theta
  d\phi^2))\right]-d\psi^2,
\end{equation}
where $\Phi(\tau,r,\theta,\phi,\psi)$ is the metric function that
describes the gauge-invariant scalar metric fluctuations. In
cartesian coordinates this perturbed line element reads
\begin{equation}\label{a7}
\left.dS_{5}^{2}\right|_{pert}=\left(\frac{\psi}{\psi_0}\right)^2 \left[F(\tau,x,y,z)e^{2\Phi}d\tau^2 - J(\tau,x,y,z)e^{-2\Phi}\delta_{ij}
dx^{i}dx^{j}\right]-d\psi^2,
\end{equation}
being now $\Phi=\Phi(\tau,x,y,z,\psi)$. For simplicity, we will use (\ref{a7}) to obtain the Einstein field equations in 5D. \\

Now let us to consider a non-massive scalar field $\varphi(x^{a})$ on the 5D spacetime (\ref{a7}). Its dynamics can be derived from the action
\begin{equation}\label{a8}
^{(5)}{\cal S}=\int \sqrt{g_5}\left[\frac{^{(5)}R}{2\kappa_5}-\frac{1}{2}g^{ab}\varphi_{,a}\varphi_{,b}\right] d^{4}x\, d\psi,
\end{equation}
where $^{(5)}R$ is the Ricci scalar, $g_5$ is the determinant of
the metric (\ref{a7}) and $\kappa_5$ is the 5D gravitational
coupling constant. As the scalar field $\varphi$ is purely kinetic
there are no interactions of this field with its environment in
5D, however, in 4D this is not the case. Clearly, the
energy-momentum tensor derived from (\ref{a8}) reads
\begin{equation}\label{a9}
^{(5)}T_{ab}=\varphi_{,a}\varphi_{,b}-\frac{1}{2}g_{ab}\,\varphi_{,c}\varphi^{,c},
\end{equation}
which is obviously symmetric. The dynamics of the scalar field $\varphi$ derived from the action (\ref{a8}) on the perturbed metric (\ref{a7})
is given by
\begin{equation}\label{a10}
\ddot{\varphi}+\left[\left(\frac{3}{2}\frac{\dot{J}}{J}-\frac{1}{2}\frac{\dot{F}}{F}\right)-4\dot{\Phi}\right]\dot{\varphi}-
\sqrt{\frac{F}{J^3}}e^{4\Phi}\vec{\nabla}(\sqrt{FJ}\,)\cdot\vec{\nabla}\varphi-\frac{F}{J}e^{4\Phi}\nabla^2\varphi+\left(\frac{\psi}{\psi_0}
\right)^2Fe^{2\Phi}\left[\left(2\overset{\star}{\Phi}-\frac{4}{\psi}\right)\overset{\star}{\varphi}-\overset{\star\star}{\varphi}\right]=0,
\end{equation}
where we denote $(\vec{\nabla}\Phi)^2 \equiv
\vec{\nabla}\Phi\cdot\vec{\nabla} \Phi$, the dot $(\cdot)$ denotes
partial derivative with respect to the conformal time $\tau$ and
the star $(\star)$ is denoting partial derivative with respect to
the extra coordinate $\psi$. The diagonal components of the 5D
field equations $^{(5)}G_{ab}=k_5\:^{(5)}T_{ab}$, on the
background (\ref{a7}), have the form
\begin{footnotesize}
\begin{eqnarray}
-\frac{3}{4}\left(\frac{\dot{J}}{J}\right)^2&+&\frac{3\dot{J}}{J}\dot{\Phi}-3\dot{\Phi}^2+\left[\frac{3F}{\psi_0^2}\right.\nonumber
\\ &-& \left.
3F\left(\frac{\psi}{\psi_0}\right)^2(\overset{\star\star}{\Phi}-2\overset{\star}{\Phi}^2)-\frac{12F\psi}{\psi_0^2}\overset{\star}{\Phi}
\right]e^{2\Phi}-\left[\frac{3F}{4J^3}(\vec{\nabla} J)^2
+\frac{F}{J^2}(\vec{\nabla}
J\cdot\vec{\nabla}\Phi)-\frac{F}{J}(\vec{\nabla}\Phi)^2+\frac{2F}{J}\nabla^2\Phi
-\frac{F}{J^2}\nabla^2J\right]e^{4\Phi}\nonumber\\
& =& -
\kappa_{5}\left[\frac{1}{2}\dot{\varphi}^2+\frac{1}{2}\left(\frac{F}{J}\right)e^{4\Phi}(\vec{\nabla}\varphi)^2+\frac{1}{2}\left(\frac{\psi}{\psi_0}
\right)^2
Fe^{2\Phi}\overset{\star}{\varphi}^2\right], \label{a11}\\
\frac{3}{4}\left(\frac{\vec{\nabla}
J}{J}\right)^2&+&\frac{1}{2}\left(\frac{\vec{\nabla}
F}{F}\right)^2-\left(\frac{\vec{\nabla} F\cdot\vec{\nabla}
J}{2FJ}\right)- \frac{\nabla^2
J}{J}-\frac{{\nabla}^2F}{F}-\frac{\vec{\nabla}
F\cdot\vec{\nabla}\Phi}{F}-(\vec{\nabla}\Phi)^2\nonumber
\\
&+&\left[-\left(\frac{9J}{\psi_0^2}\right)-
6J\left(\frac{\psi}{\psi_0}\right)^2\overset{\star}{\Phi}^2+\frac{12J\psi}{\psi_0^2}\overset{\star}{\Phi}+3J\left(\frac{\psi}{\psi_0}\right)^2
\overset{\star\star}{\Phi}
\right]e^{-2\Phi}\nonumber \\
&+&\left[\left(\frac{3J\dot{F}}{F^2}-\frac{12\dot{J}}{F}\right)\dot{\Phi}-\frac{3\dot{J}^2}{4FJ}-\frac{3\dot{F}\dot{J}}{2F^2}+\frac{3\ddot{J}}{F}
+\frac{15J}{F}
\dot{\Phi}^2-\frac{6J}{F}\ddot{\Phi}\right]e^{-4\Phi}\nonumber \\
&=&
-\kappa_{5}\left[\frac{3J}{2F}e^{-4\Phi}\dot{\varphi}^2-\frac{1}{2}(\vec{\nabla}\varphi)^2-\frac{3J}{2}
\left(\frac{\psi}{\psi_0}\right)^2e^{-2\Phi}\overset{\star}{\varphi}^2\right],\label{a12} \\
-\frac{6}{\psi^2}+\frac{6}{\psi}\overset{\star}{\Phi}&+&\left(\frac{\psi_0}{\psi}\right)^2
\left[\frac{(\vec{\nabla} F)^2}{4JF^2}+\frac{3(\vec{\nabla}
J)^2}{4J^3} -\frac{ \vec{\nabla} F\cdot\vec{\nabla}
J}{4FJ^2}-\frac{\nabla^2J}{J^2}-\frac{\nabla^2
F}{2FJ}-\frac{1}{2}\left(\frac{\vec{\nabla}
F}{FJ}-\frac{\vec{\nabla} J}{J^2}\right)
\cdot\vec{\nabla}\Phi-\frac{(\vec{\nabla}\Phi)^2}{J}+\frac{\nabla^2 \Phi}{J}\right]e^{2\Phi}\nonumber \\
&+&\left(\frac{\psi_0}{\psi}\right)^2\left[-\frac{3}{4}\frac{\dot{F}\dot{J}}{F^2
J} + \frac{3}{2}\frac{\ddot{J}}{FJ}-\frac{3}{2}
\left(\frac{5\dot{J}}{FJ}-\frac{\dot{F}}{F^2}\right)\dot{\Phi}+\frac{9}{F}\dot{\Phi}^2-\frac{3}{F}\ddot{\Phi}
\right]e^{-2\Phi}\nonumber \\
&=&
-\kappa_5\left[\frac{1}{2F}\left(\frac{\psi}{\psi_0}\right)^{-2}e^{-2\Phi}\dot{\varphi}^2-\frac{1}{2J}\left(\frac{\psi}{\psi_0}
\right)^{-2}e^{2\Phi}(\vec{\nabla}\varphi)^2+\frac{1}{2}\overset{\star}{\varphi}^2\right]\label{a13},
\end{eqnarray}
\end{footnotesize}
while the non-diagonal 5D Einstein equations are given by
\begin{small}
\begin{eqnarray}
 \frac{\dot{J}_{,i}}{J}-\frac{1}{2}\frac{\dot{J}F_{,i}}{JF}-\frac{\dot{J}J_{,i}}{J^2}+2\Phi_{,i}\dot{\Phi}-\left(\frac{\dot{J}}{J}\right)
 \Phi_{,i}-2\dot{\Phi}_{,i}+\left(\frac{F_{,i}}{F}\right)\dot{\Phi}&=& - \kappa_5 \dot{\varphi}\varphi_{,i}\, ,\label{a14}\\
 6\overset{\star}{\Phi}\dot{\Phi}-3\left(\frac{\dot{J}}{J}\right)\overset{\star}{\Phi}-3\overset{\cdot\,\star}{\Phi}&=&
 -\kappa_5\dot{\varphi}\overset{\star}{\varphi},\label{a15}\\
2\Phi_{,i}\overset{\star}{\Phi}+\left(\frac{F_{,i}}{F}\right)\overset{\star}{\Phi}-\overset{\star}{\Phi}_{,i} &=&
-\kappa_{5} \varphi_{,i}\overset{\star}{\varphi},\\
2\Phi_{,i}\Phi_{,j}+\frac{F_{,i}}{F}\Phi_{,j}+\frac{F_{,j}}{F}\Phi_{,i}+\frac{1}{2}\frac{F_{,ij}}{F}-\frac{1}{4}
\frac{F_{,i}F_{,j}}{F^2}-\frac{1}{4}\frac{F_{,i}J_{,j}}{FJ}-\frac{1}{4}\frac{F_{,j}J_{,i}}{FJ}+\frac{J_{,ij}}{2J}
-\frac{3}{4}\frac{J_{,i}J_{,j}}{J^2} &=& - \kappa_{5}\varphi_{,i}\varphi_{,j}\quad\mbox{for}\quad i\neq j.\label{a16}
\end{eqnarray}
\end{small}

\section{5D gauge invariant metric fluctuations on Cosmological scales}

On cosmological scales the following condition is satisfied:
\begin{equation}\label{a17}
\frac{G\zeta\psi_0}{2a(\tau)r_{H}}\ll 1,
\end{equation}
where $r_{H}$ denotes the value of the radial coordinate at the
horizon entry. This means that on cosmological scales the
functions $F(\tau,r)$ and $J(\tau,r)$ become independent of
spatial coordinates [see eqs. in (\ref{a4})], so that
\begin{equation}
\left.F(\tau,r)\right|_{{G\xi\psi_0\over 2 a(\tau) r_H} \ll 1}
\rightarrow a^2(\tau), \qquad
\left.J(\tau,r)\right|_{{G\xi\psi_0\over 2 a(\tau) r_H} \ll 1}
\rightarrow a^2(\tau),
\end{equation}
and the metric (\ref{a3}) describes an universe which is nearly 3D
spatially homogeneous and  isotropic.

The perturbed field equations (\ref{a11})-(\ref{a13}) with the
condition (\ref{a17}), give us the diagonal Einstein equations on
cosmological scales

\begin{eqnarray}
 3{\cal H}^2-6{\cal H}\dot{\Phi}+3\dot{\Phi}^2-[(\vec{\nabla}\Phi)^2-2\nabla^2\Phi]e^{4\Phi}-3a^2\left[\frac{1}{\psi_0^2}
 -\left(\frac{\psi}{\psi_0}\right)^2\left(\overset{\star\star}{\Phi}-2\overset{\star}{\Phi}^2\right)
 - \frac{4\psi}{\psi_0^2}\overset{\star}{\Phi}\right]e^{2\Phi}\nonumber \\
=\frac{\kappa_5}{2}\left[\dot{\varphi}^2+e^{4\Phi}(\vec{\nabla}\varphi)^2+\left(\frac{\psi}{\psi_0}\right)^2a^2e^{2\Phi}\overset{\star}{\varphi}^2
\right],\label{a18}\\
-{\cal H}^2-2\dot{\cal H}+2\ddot{\Phi}+6{\cal
H}\dot{\Phi}-5\dot{\Phi}^2+\frac{1}{3}e^{4\Phi}(\vec{\nabla}\Phi)^2
+a^2\left[\frac{3}{\psi_0^2}+2\left(\frac{\psi}{\psi_0}\right)^2\overset{\star}{\Phi}^2-\frac{4\psi}{\psi_0^2}\overset{\star}{\Phi}
-\left(\frac{\psi}{\psi_0}\right)^2\overset{\star\star}{\Phi}\right]e^{2\Phi}\nonumber\\
=\frac{\kappa_5}{2}\left[\dot{\varphi}^2-\frac{1}{3}e^{4\Phi}(\vec{\nabla}\varphi)^2-\left(\frac{\psi}{\psi_0}\right)^2a^2e^{2\Phi}
\overset{\star}{\varphi}^2\right], \label{a19}\\
-3({\cal H}^2+\dot{\cal H})+3(\ddot{\Phi}+4{\cal
H}\dot{\Phi}-3\dot{\Phi}^2)+\left[(\vec{\nabla}\Phi)^2-\nabla^2\Phi\right]e^{4\Phi}
+\frac{6a^2}{\psi_0^2}(1-\psi\overset{\star}{\Phi})e^{2\Phi}\nonumber\\
=\frac{\kappa_5}{2}\left[\dot{\varphi}^2-e^{4\Phi}(\nabla\varphi)^2+\left( \frac{\psi}{\psi_0}\right)^2a^2e^{2\Phi}\overset{\star}{\varphi}^2
\right], \label{a20}
\end{eqnarray}
whereas the non-diagonal equations (\ref{a14}) to (\ref{a16}) become
\begin{eqnarray}
-2\Phi_{,i}\dot{\Phi}+2{\cal H}\Phi_{,i}+2\dot{\Phi}_{,i}&=&\kappa_{5}\dot{\varphi}\varphi_{,i},\label{a21}\\
-6\overset{\star}{\Phi}\dot{\Phi}+6{\cal H}\overset{\star}{\Phi}+3\overset{\cdot\,\star}{\Phi}&=&\kappa_5\dot{\varphi}\overset{\star}{\varphi},\label{a22}\\
-2\Phi_{,i}\overset{\star}{\Phi}+\overset{\star}{\Phi}_{,i}&=&\kappa_{5}\varphi_{,i}\overset{\star}{\varphi},\label{a23}\\
-2\Phi_{,i}\Phi_{,j}&=&\kappa_{5}\varphi_{,i}\varphi_{,j}\quad\mbox{for}\quad
i\neq j.\label{a24}
\end{eqnarray}
Here, ${\cal H}=\dot{a}(\tau)/a(\tau)$ is the Hubble parameter in
conformal time. From the linear combination [eq (\ref{a20})- eq.
(\ref{a19}) + (1/4) $\times$ eq. (\ref{a18})] results the
expression
\begin{eqnarray}
-\frac{5}{4}{\cal H}^2-\dot{\cal H}+\ddot{\Phi}+\frac{9}{2}{\cal
H}\dot{\Phi}-\frac{13}{4}\dot{\Phi}^2+\frac{5}{12}
(\vec{\nabla}\Phi)^2e^{4\Phi}-\frac{1}{2}(\nabla^2\Phi)e^{4\Phi}+a^2\left[\frac{9}{4\psi_0^2}-\frac{7}{2}
\left(\frac{\psi}{\psi_0}\right)^2\left(\overset{\star}{\Phi}^2-\frac{1}{2}\overset{\star\star}{\Phi}
\right)+\frac{\psi}{\psi_0^2}\overset{\star}{\Phi}\right]e^{2\Phi}\nonumber\\
=\frac{\kappa_5}{2}\left[\frac{1}{4}\dot{\varphi}^2-\frac{5}{12}e^{4\Phi}(\vec{\nabla}\varphi)^2+\frac{9}{4}
\left(\frac{\psi}{\psi_0}\right)^2a^2e^{2\Phi}\overset{\star}{\varphi}^2\right].\quad
\label{a25}
\end{eqnarray}
This equation describes the dynamics of the 5D scalar metric
fluctuations $\Phi(\tau,x,y,z,\psi)$ in terms of the 5D scalar
field $\varphi(\tau,x,y,z,\psi)$, which in 4D will be formally
identified with the inflaton field. Finally, equation (\ref{a10})
for $\varphi(\tau,x,y,z,\psi)$ yields
\begin{equation}\label{a26}
\ddot{\varphi}+(2{\cal H}-4\dot{\Phi})\dot{\varphi}-e^{4\Phi}\nabla^{2}\varphi+\left(\frac{\psi}{\psi_0}\right)^2a^2e^{2\Phi}
\left[\left(2\overset{\star}{\Phi}-\frac{4}{\psi}\right)\overset{\star}{\varphi}-\overset{\star\star}{\varphi}\right]=0,
\end{equation}
which is the equation that describes the 5D dynamics of the scalar
field $\varphi$ on cosmological scales in the presence of the
scalar metric fluctuations.

\section{Weak field limit of 5D metric fluctuations}

In the previous section we have derived the 5D field equations
describing the dynamics of the 5D gauge-invariant scalar metric
fluctuations $\Phi$, in a non-perturbative manner. The amplitudes
of these metric fluctuations have no restrictions, neither on
cosmological nor astrophysical scales. However, as is well known,
such amplitudes are small on cosmological scales and specially
during an inflationary stage of the early universe. Due to this
fact we shall consider small amplitudes, in order to study the
spectrum of these scalar metric fluctuations during inflation on
large-scales. We shall consider the weak field limit
approximation: $e^{\pm 2\Phi}\simeq 1\pm 2\Phi$.\\

Let us start with the weak field limit of equation (\ref{a26}).
Linearizing this equation respect to $\Phi$ we obtain
\begin{equation}\label{a27}
\ddot{\varphi}-(2{\cal H}-4\dot{\Phi})\dot{\varphi}-(1+4\Phi)\nabla^{2}\varphi+\left(\frac{\psi}{\psi_0}\right)^2a^2
\left[\left(2\overset{\star}{\Phi}-\frac{4}{\psi}\right)\overset{\star}{\varphi}-(1+2\Phi)\overset{\star\star}{\varphi}-\frac{8}{\psi}
\Phi\overset{\star}{\varphi}\right]=0.
\end{equation}
Now, implementing a semi-classical approximation for the scalar
field: $\varphi(\tau,x,y,z,\psi)=\varphi_{b}(\tau,\psi)
+\delta\varphi(\tau,x,y,z,\psi)$, in equation (\ref{a27}), we
obtain separately the dynamics for both, the background part of
the field $\varphi_b$ and for the quantum fluctuations
$\delta\varphi$
\begin{eqnarray}
&&\ddot{\varphi}_b+2{\cal H}\dot{\varphi}_b-\left(\frac{\psi}{\psi_0}\right)^2a^2\left[\frac{4}{\psi}\overset{\star}{\varphi}_b
+\overset{\star\star}{\varphi}_b\right]=0,\label{a28}\\
&& \ddot{\delta\varphi}+2{\cal H}\dot{\delta\varphi}-\nabla^{2}\delta\varphi-\left(\frac{\psi}{\psi_0}\right)^2a^2
\left[\frac{4}{\psi}\overset{\star}{\delta\varphi}+\overset{\star\star}{\delta\varphi}\right]-4\dot{\varphi}_{b}\dot{\Phi}
+\left(\frac{\psi}{\psi_0}\right)^{2}a^2\left[2\overset{\star}{\varphi}_{b}\overset{\star}{\Phi}-\left(\frac{8}{\psi}\overset{\star}{\varphi}_{b}
+2\overset{\star\star}{\varphi}_b\right)\Phi\right]=0. \label{a29}
\end{eqnarray}
Similarly, linearizing with respect to $\Phi$ equations
(\ref{a18}) to (\ref{a20}), we obtain for the background field
$\varphi_b$
\begin{eqnarray}
3{\cal H}^2-\frac{3a^2}{\psi_0^2}=\frac{\kappa_5}{2}\left[\dot{\varphi}_{b}^2+\left(\frac{\psi}{\psi_0}\right)^2 a^2 \overset{\star}{\varphi}_b^2
\right], \label{a30}\\
-{\cal H}^{2}-2\dot{{\cal H}}+\frac{3a^{2}}{\psi_{0}^{2}} = \frac{\kappa_5}{2}\left[\dot{\varphi}_b^2-\left(\frac{\psi}{\psi_0}
\right)^2a^2\overset{\star}{\varphi}_b^2  \right], \label{a31}\\
-3({\cal H}^2+\dot{\cal H})+\frac{6a^2}{\psi_0^2}=\frac{\kappa_5}{2}\left[\dot{\varphi}_b^2+\left(\frac{\psi}{\psi_0}\right)^2a^2
\overset{\star}{\varphi}_b^2\right], \label{a32}
\end{eqnarray}
while for that for the field $\delta\varphi$, we have
\begin{eqnarray}
-6{\cal H}\dot{\Phi}+2\nabla^2\Phi+3a^2\left[\left(\frac{\psi}{\psi_0}\right)^2\overset{\star\star}{\Phi}+\frac{4\psi}{\psi_0^2}\overset{\star}{\Phi}
-\frac{2}{\psi_0^2}\Phi\right]=\kappa_5\left[\dot{\varphi}_b\delta\dot{\varphi}+\left(\frac{\psi}{\psi_0}\right)^2a^2
\left(\overset{\star}{\varphi}_b\overset{\star}{\delta\varphi}+\overset{\star}{\varphi}_b^2\Phi\right)\right], \label{a33}\\
2\ddot{\Phi}+6{\cal H}\dot{\Phi}+a^2\left[\frac{6}{\psi_0^2}\Phi-\left(\frac{\psi}{\psi_0}\right)^2\overset{\star\star}{\Phi}
-\frac{4\psi}{\psi_0^2}\overset{\star}{\Phi}\right]=\kappa_5\left[\dot{\varphi}_b\dot{\delta\varphi}-\left(\frac{\psi}{\psi_0}\right)^2a^2
\left(\overset{\star}{\varphi}_b\overset{\star}{\delta\varphi}+\overset{\star}{\varphi}_b^2\Phi\right)\right], \label{a34}\\
3(\ddot{\Phi}+4{\cal H}\dot{\Phi})-\nabla^2\Phi+\frac{6a^2}{\psi_0^2}\left(2\Phi-\psi\overset{\star}{\Phi}\right)=\kappa_5
\left[\dot{\varphi}_b\dot{\delta\varphi}+\left(\frac{\psi}{\psi_0}\right)^2a^2\left(\overset{\star}{\varphi}_b\overset{\star}{\delta\varphi}
+\overset{\star}{\varphi}_b^2\Phi\right)
\right]. \label{a35}
\end{eqnarray}
On the other hand, linearizing the non-diagonal field Einstein
equations (\ref{a21}) to (\ref{a24}), we obtain
\begin{eqnarray}
2\left({\cal H}\Phi_{,i}+\dot{\Phi}_{,i}\right)&=&\kappa_5\dot{\varphi}_b\delta\varphi_{,i}\label{36}\\
6{\cal H}\overset{\star}{\Phi}+3\overset{\cdot\star}{\Phi}&=&\kappa_5
\left(\dot{\varphi}_b\overset{\star}{\delta\varphi}+\dot{\delta\varphi}\overset{\star}{\varphi}_b\right) \label{a37}\\
\overset{\star}{\Phi}_{,i}&=&\kappa_5\overset{\star}{\varphi}_b\delta\varphi_{,i}.\label{a38}
\end{eqnarray}
Equation (\ref{a25}) after being linearized can be written as
\begin{equation}\label{a39}
-\frac{5}{4}{\cal H}^2-\dot{\cal H}+\frac{9a^2}{4\psi_0^2}=\frac{\kappa_5}{8}\left[\dot{\varphi}_b^2+9\left(\frac{\psi}{\psi_0}\right)^2a^2
\overset{\star}{\varphi}_b^2\right],
\end{equation}
while that the dynamics of the field $\delta\varphi$, is given by
\begin{equation}\label{a40}
\ddot{\Phi}+\frac{9}{2}{\cal H}\dot{\Phi}-\frac{1}{2}\nabla^2\Phi+a^2\left[\frac{9}{2\psi_0^2}\Phi+\frac{\psi}{\psi_0^2}\overset{\star}{\Phi}
+\frac{7}{4}\left(\frac{\psi}{\psi_0}\right)^2\overset{\star\star}{\Phi}\right]=\frac{\kappa_5}{4}\left[\dot{\varphi}_b\dot{\delta\varphi}
+9\left(\frac{\psi}{\psi_0}\right)^2a^2\left(\overset{\star}{\varphi}_b\overset{\star}{\delta\varphi}+\overset{\star}{\varphi}_b^2\Phi\right)\right].
\end{equation}
This last equation describes the 5D dynamics of the metric
fluctuations $\Phi$ in terms of the scalar field fluctuations
$\delta\varphi$. Clearly, this equation can be also derived as a
linear combination of equations (\ref{a33})-(\ref{a35}).

\section{Inducing the 4D dynamics for the non-perturbative gauge invariant scalar metric fluctuations }

Since we know the 5D field equations of motion for both, the
scalar field $\varphi$ and the non-perturbative scalar metric
fluctuations $\Phi$, we are in a position to derive their
respective dynamics on our 4D universe. With this in mind, we
shall assume that the 5D spacetime can be foliated by a family of
hypersurfaces $\Sigma:\psi=constant$. Our 4D universe will be here
represented by a generic hypersurface $\Sigma_{0}:\psi=\psi_0$.
Thus, on every leaf member of the family, the line element induced
by (\ref{a3}) has the form
\begin{equation}\label{b1}
ds_4^2=F(\tau,r)d\tau^2-J(\tau,r)[dr^2+r^2(d\theta^2+sin^2\theta d\phi^2)],
\end{equation}
with
\begin{equation}\label{b2}
F(\tau,r)=a^2(\tau)\left[1-\frac{Gm}{2ar}\right]^2\left[1+\frac{Gm}{2ar}\right]^{-2},\quad J(\tau,r)=a^{2}(\tau)\left[1+\frac{Gm}{2ar}\right]^{4},
\end{equation}
where $m=\zeta\psi_0$ is the physical mass of a primordial
black-hole which evolves during inflation and has a Schwarzschild
radius $R_H = 2 G m$. It is clear that the condition (\ref{a17})
delimits the cosmological scales. On the 4D hypersurface, this
condition takes the form
\begin{equation}\label{b3}
\frac{Gm}{2ar}\ll 1.
\end{equation}
The background metric in (\ref{b1}) describes on cosmological
scales a de Sitter expansion of the early universe characterized
by an equation of state $p_{b}=-\rho_{b}=-3/(8\pi G \psi^2_0)$,
where $\rho_b$ and $p_b$ are the background energy density and the
pressure. The 5D perturbed line element in cartesian coordinates
(\ref{a7}), induces on the hypersurface $\Sigma_0$ the effective
4D line element
\begin{equation}\label{b4}
\left.ds_{4}^2\right|_{pert}=F(\tau,\bar{x})e^{2\Omega(\tau,\bar{x})}d\tau^2-J(\tau,\bar{x})e^{-2\Omega(\tau,\bar{x})}\delta_{ij}dx^{i}dx^{j},
\end{equation}
where $\Omega(\tau,\bar{x})\equiv\Phi(\tau,x,y,z,\psi_0)$
describes the 4D scalar metric fluctuations induced on $\Sigma_0$.
Equations (\ref{a18})-(\ref{a20}) induce on our 4D spacetime
$\Sigma_0$, the field equations
\begin{eqnarray}
3{\cal H}^2-6{\cal
H}\dot{\Omega}+3\dot{\Omega}^2-[(\vec{\nabla}\Omega)^2-2\nabla^2\Omega]e^{4\Omega}-3a^2e^{2\Omega}\left.\left[\frac{1}{\psi_0^2}
-\left(\frac{\psi}{\psi_0}\right)^2\left(\overset{\star\star}{\Phi}-2\overset{\star}{\Phi}^2\right)-\frac{4\psi}{\psi_0^2}\overset{\star}{\Phi}
\right]\right|_{\psi=\psi_0}\nonumber \\
= 4\pi
G\left[\dot{\bar{\varphi}}^2+e^{4\Omega}(\vec{\nabla}\bar{\varphi})^2+a^2e^{2\Omega}\left.\left(\frac{\psi^2}{\psi_0^2}\overset{\star}{\varphi}^2
\right)\right|_{\psi=\psi_0}\right],\label{b5}\\
-{\cal H}^2-2\dot{\cal H}+2\ddot{\Omega}+6{\cal
H}\dot{\Omega}-5\dot{\Omega}^2+\frac{1}{3}e^{4\Omega}(\vec{\nabla}\Omega)^2+a^2e^{2\Omega}\left.
\left[\frac{3}{\psi_0^2}+2\left(\frac{\psi}{\psi_0}\right)^2\overset{\star}{\Phi}^2-4\frac{\psi}{\psi_0^2}\overset{\star}{\Phi}
-\left(\frac{\psi}{\psi_0}\right)^2\overset{\star\star}{\Phi}\right]\right|_{\psi=\psi_0}\nonumber \\
= 4\pi
G\left[\dot{\bar{\varphi}}^2-\frac{1}{3}e^{4\Omega}(\vec{\nabla}\bar{\varphi})^2-a^2e^{2\Omega}\left.
\left(\frac{\psi^2}{\psi_0^2}\overset{\star}{\varphi}^2\right)\right|_{\psi=\psi_0}\right], \label{b6}\\
-3({\cal H}^2+\dot{\cal H})+3(\ddot{\Omega}+4{\cal
H}\dot{\Omega}^2-3\dot{\Omega}^2)+[(\vec{\nabla}\Omega)^2
-\nabla^2\Omega]e^{4\Omega}+6a^2e^{2\Omega}\left.\left[\frac{1}{\psi_0^2}-\frac{\psi}{\psi_0^2}\overset{\star}{\Phi}\right]\right|_{\psi=\psi_0}
\nonumber\\
= 4\pi
G\left[\dot{\bar{\varphi}}^2-e^{4\Omega}(\vec{\nabla}\bar{\varphi})^2+a^2e^{2\Omega}\left.
\left(\frac{\psi^2}{\psi_0^2}\overset{\star}{\varphi}^2\right)\right|_{\psi=\psi_0}\right],
\label{b7}
\end{eqnarray}
where
$\bar{\varphi}(\tau,x,y,z)=\varphi(\tau,x,y,z,\psi)|_{\psi=\psi_0}$
is the massive scalar field induced on $\Sigma_0$. The dynamics of
the induced scalar metric fluctuations $\Omega$, derived from
(\ref{a25}), is given by
\begin{eqnarray}
&-&\frac{5}{4}{\cal H}^2-\dot{\cal
H}+\ddot{\Omega}+\frac{9}{2}{\cal H}\dot{\Omega}
-\frac{13}{4}\dot{\Omega}^2+\frac{5}{12}(\vec{\nabla}\Omega)^2e^{4\Omega}-\frac{1}{2}(\nabla^2\Omega)e^{4\Omega}\nonumber \\
&+&a^2e^{2\Omega}
\left.\left[\frac{9}{4\psi_0^2}-\frac{7}{2}\left(\frac{\psi}{\psi_0}
\right)^2\overset{\star}{\Phi}^2+\frac{\psi}{\psi_0^2}\overset{\star}{\Phi}+\frac{7}{4}\left(\frac{\psi}{\psi_0}
\right)^2\overset{\star\star}{\Phi}\right]\right|_{\psi=\psi_0}\nonumber\\
&=&\frac{8\pi
G}{2}\left[\frac{1}{4}\dot{\bar{\varphi}}^2-\frac{5}{12}e^{4\Omega}(\vec{\nabla}\bar{\varphi})^2
+\frac{9}{4}a^2e^{2\Omega}\left.\left(\frac{\psi^2}{\psi_0^2}\overset{\star}{\varphi}^2\right)\right|_{\psi=\psi_0}\right],
\label{b8}
\end{eqnarray}
and the induced dynamics of $\bar{\varphi}$ on the hypersurface 4D
is given by the equation
\begin{equation}\label{b9}
\ddot{\bar{\varphi}}+\left[2{\cal H}-4\dot{\Omega}\right]\dot{\bar{\varphi}}-e^{4\Omega}\nabla^2{\bar{\varphi}}
+ae^{2\Omega}\left.\left[\left(2\overset{\star}{\Phi}-\frac{4}{\psi}\right)\overset{\star}{\varphi}-\overset{\star\star}{\varphi}
\right]\right|_{\psi=\psi_0}=0,
\end{equation}
which is derived by evaluating (\ref{a26}) on $\Sigma_0$.

The 5D action (\ref{a8}) induces on our 4D spacetime the effective action
\begin{equation}\label{b10}
^{(4)}{\cal S}_{eff}=\int d^{4}x\sqrt{g_4}\left[\frac{^{(4)}R}{16\pi G}-\frac{1}{2}g^{\mu\nu}\bar{\varphi}_{,\mu}\bar{\varphi}_{,\nu}
+V(\bar{\varphi})\right],
\end{equation}
where $g_4$ is the determinant of the 4D induced metric, which for the background reads $g_{4}^{(b)}=FJ^{3}$ while for
the perturbed metric $g_{4}^{(p)}=FJ^3e^{-4\Omega}$. The 4D Ricci scalar curvature $^{(4)}R$ is given by
\begin{equation}\label{b11}
^{(4)}R=\frac{2}{a^2}\left[3({\cal H}^2+\dot{\cal H}-4{\cal H}\dot{\Omega}-\ddot{\Omega}+3\dot{\Omega}^2)e^{-2\Omega}+(\nabla^2\Omega
-(\nabla\Omega)^2)e^{2\Omega}\right],
\end{equation}
and the induced 4D effective potential $V$ has the form
\begin{equation}\label{b12}
V(\bar{\varphi})=-\frac{1}{2}g^{\psi\psi}\left.\left(\frac{\partial\varphi}{\partial\psi}\right)^2\right|_{\psi=\psi_0}.
\end{equation}
In our analysis the fields $\Omega$ and $\bar{\varphi}$ are
semi-classical fields, so they are constituted by a classical part
plus a quantum part. To study the dynamics of the former, a
standard quantization procedure will be implemented. To do it, we
shall impose the commutation relations
\begin{equation}\label{b13}
\left[\bar{\varphi}(\tau,\bar{x}),\Pi_{(\bar{\varphi})}^{0}(\tau,\bar{x}^{\prime})\right]=i\delta^{(3)}(\bar{x}
-\bar{x}^{\prime}),\quad \left[\Omega(\tau,\bar{x}),\Pi_{(\Omega)}^{0}(\tau,\bar{x}^{\prime})\right]=i\delta^{(3)}(\bar{x}-\bar{x}^{\prime}),
\end{equation}
where $\bar{x}$ is denoting the 3D vector position in cartesian
coordinates. Due to the fact that the conjugate momentum to
$\bar{\varphi}$ and $\Omega$ are, respectively, given by
$\Pi_{(\bar{\varphi})}^{0}=\sqrt{-g_4}F^{-1}e^{-2\Omega}\dot{\bar{\varphi}}$
and $\Pi_{(\Omega)}^{0}=[12/(16\pi G)]a^{-2}(3\dot{\Omega}-2{\cal
H})e^{-2\Omega}\sqrt{-g_4}$, the expressions (\ref{b13}) yield
\begin{equation}\label{b14}
\left[\bar{\varphi}(\tau,\bar{x}),\dot{\bar{\varphi}}(\tau,\bar{x}^{\prime})\right]=\frac{ie^{2\Omega}}{F\sqrt{-g_4}}
\delta^{(3)}(\bar{x}-\bar{x}^{\prime}),\quad
\left[\Omega(\tau,\bar{x}),\dot{\Omega}(\tau,\bar{x}^{\prime})\right]
=i\frac{4\pi G
a^2e^{2\Omega}}{9\sqrt{-g_4}}\delta^{(3)}(\bar{x}-\bar{x}^{\prime}).
\end{equation}

\section{Induced 4D dynamics  for the inflaton and  gauge invariant metric fluctuations in the weak field limit}

In the section IV we have obtained the weak field limit of the 5D
field equations for both fields: the inflaton field $\varphi$ and
the scale-invariant metric fluctuation field $\Phi$. A similar
mechanism can be applied in the case of the effective 4D field
equations of the section V. Thus, let us use the semiclassical
approximation: $\bar{\varphi}(\tau,\bar{r})
=\bar{\varphi}_{b}(\tau)+\delta\bar{\varphi}(\tau,\vec{r})$, where
$\bar{\varphi}_b(\tau)\equiv
\left.\bar{\varphi}_b(\tau,\psi)\right|_{\psi=\psi_0}$ is the
background 4D inflaton field and
$\delta\bar{\varphi}(\tau,\vec{r})\equiv
\left.\delta\bar{\varphi}(\tau,\vec{r},\psi)\right|_{\psi=\psi_0}$
stands for the 4D inflaton field quantum fluctuations. Taking this
into account, the evaluation of equation (\ref{a28}) on $\Sigma_0$
yields
\begin{equation}\label{b15}
\ddot{\bar{\varphi}}_b+2{\cal H}\dot{\bar{\varphi}}_b+a^2m^2\bar{\varphi}_b=0,
\end{equation}
where we have used the relation:
$[(4/\psi)\overset{\star}{\varphi}_b+\overset{\star\star}{\varphi}_b]|_{\psi=\psi_0}
=-m^2\bar{\varphi}_b$, such that $m$ is a separation constant.
Now, according to equation (\ref{b5}), the background field
$\bar{\varphi}$ must obey the Friedmann-like equation
\begin{equation}\label{b16}
\left(\frac{\partial\bar{\varphi}_b}{\partial\tau}\right)^2
+a^2\left(\frac{\partial\varphi_b}{\partial\psi}\right)^2_{\psi=\psi_0}=0.
\end{equation}
A particular solution of (\ref{b15}), which also is satisfied when
inflation begins, are the slow rolling conditions:
$\partial\bar{\varphi}_b/\partial\tau=0$, where necessarily $m=0$
\cite{cop}. Using this solution in (\ref{b16}), it yields
\begin{equation}\label{b17}
\bar{\varphi}_{b}=0,
\end{equation}
which means that all the energy density on the 4D hypersurface is
induced geometrically by the foliation $\psi=\psi_0=1/H$, because
the background energy density related to the background inflaton field is null. \\

\begin{figure*}
\includegraphics[height=15cm]{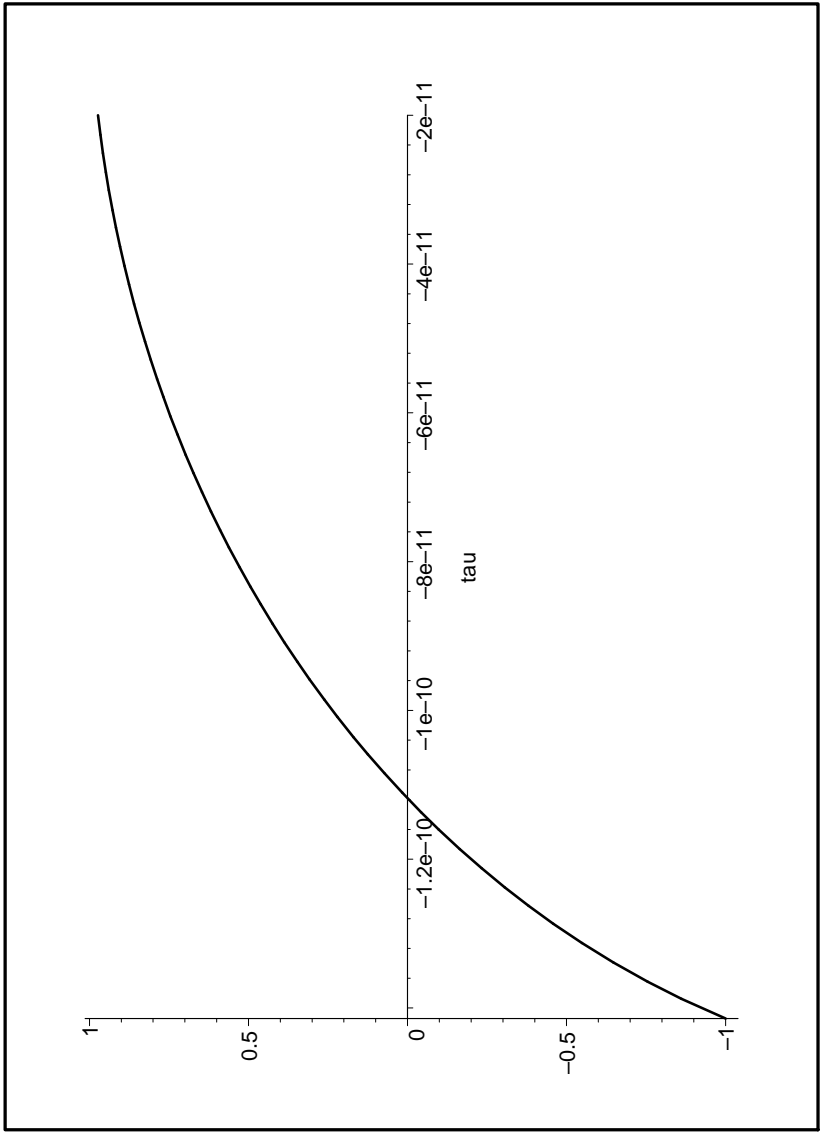}\caption{\label{f1} Evolution of $n_s(\tau)$ during inflation for $\lambda=10^{-10}\, G^{-1/2}$.
The universe becomes
scale invariant with the growth of the universe. }
\end{figure*}

On the other hand, by using (\ref{a37}), (\ref{a38}) and
(\ref{a40}), we find after some algebra that the scalar metric
fluctuations $\Omega$ on $\Sigma_0$ obey the equation
\begin{equation}\label{b18}
\ddot{\Omega}+\frac{9}{2}{\cal
H}\dot{\Omega}-\frac{1}{2}\nabla^2\Omega+\left[\frac{9}{2}a ^2
H^2+\lambda^2\right]\Omega=0.
\end{equation}
Here, we have used
$\lbrace(\psi/\psi_0^2-9/\psi)\overset{\star}{\Phi}+[(7/4)(\psi/\psi_0)^2-9/4]\overset{\star\star}{\Phi}\rbrace_{\psi=\psi_0}=\lambda^2\Omega$,
where $\lambda$ is a separation constant with mass units. Now it
can be shown that equation (\ref{36}), evaluated on $\Sigma_0$,
leads to the condition $\dot{\Omega}=-{\cal H}\Omega$. Using this
last condition in equation (\ref{b18}), we obtain
\begin{equation}\label{b19}
\ddot{\Omega}+2{\cal
H}\dot{\Omega}-\frac{1}{2}\nabla^2\Omega+\left[a^2H^2+\lambda^2\right]\Omega=0.
\end{equation}
If we introduce the auxiliary field $\chi(\tau,\bar{r})$, through
the formula
\begin{equation}\label{b20}
\Omega(\tau,\bar{r})=e^{-\int{\cal H}(\tau)d\tau}\chi(\tau,\bar{r})
\end{equation}
the equation (\ref{b19}), becomes
\begin{equation}\label{b21}
\ddot{\chi}-\frac{1}{2}\nabla^2\chi+\left(\lambda^2-\dot{\cal
H}\right)\chi=0.
\end{equation}
Now, in order to quantize the field $\chi$, we shall expand it in
Fourier modes
\begin{equation}\label{b22}
\chi(\tau,\bar{r})=\frac{1}{(2\pi)^{3/2}}\int d^3k\left[a_ke^{i\bar{k}\cdot\bar{r}}\xi_k(\tau)+a^{\dagger}_{k}\xi_{k}^{*}(\tau)\right],
\end{equation}
where the annihilation and creation operators $a_k$ and
$a_{k}^{\dagger}$ satisfy the commutation algebra
\begin{equation}\label{b23}
[a_k,a_{k^{\prime}}^{\dagger}]=\delta^{(3)}(\bar{k}-\bar{k}^{\prime}),\qquad [a_k,a_{k^{\prime}}]=[a_{k}^{\dagger},a_{k^{\prime}}^{\dagger}]=0.
\end{equation}
Inserting (\ref{b22}) in (\ref{b21}) and using (\ref{b23}), we
find
\begin{equation}\label{b24}
\ddot{\xi}_k+\left(k_{eff}^2-\frac{2}{\tau^2}+\lambda^2\right)\xi_k=0,
\end{equation}
where $k_{eff}^2=k^2/2$. Using (\ref{b22}) and (\ref{b23}) in the
second expression of (\ref{b14}), the modes $\xi_k$ must satisfy
in the UV-sector the normalization condition
\begin{equation}\label{b25}
\xi_k\dot{\xi}_k^{*}-\xi_k^{*}\dot{\xi}_k=i\frac{4\pi G}{9a_0^2}.
\end{equation}
Thus, choosing the Bunch-Davies vacuum condition, the normalized solution of (\ref{b24}) reads
\begin{equation}\label{b26}
\xi_k(\tau)\simeq\frac{i\pi}{3a_0}\sqrt{G}{\cal
H}_{\nu}^{(2)}[z(\tau)],
\end{equation}
where ${\cal H}_{\nu}^{(2)}[z(\tau)]$ is the second kind Hankel
function, $\nu=(1/2)\sqrt{1+4\beta}$ and $z(\tau)=k_{eff} \tau$.
Notice that $\beta=2-\lambda^2\tau^2>0$, such that the conformal
time at the beginning of inflation $\tau_0= -{\sqrt{2}\over
\lambda}$ is defined such that $\beta_0=2-\lambda^2\tau^2_0=0$ and
the conformal time at the end of inflation complies with
$\beta_e=2-\lambda^2\tau^2_e \simeq 2$. Using the definition of
$\beta$, it is easy to show that the parameter $\nu\simeq 3/2$ for
$\tau^2_e \ll 2/\lambda^2$, which assures that the spectral index
$1>n_s > 0.96$\cite{rpp}. Using the fact that $1-n_s=3-2\nu$, we
obtain that the range of acceptable values for $\tau_e$ is
\begin{equation}
0 < (-\tau_e) < \frac{0.245}{\lambda}. \label{ta}
\end{equation}
Figure (\ref{f1}) shows the evolution of $n_s(\tau)$ during
inflation for $\lambda=10^{-10}\, G^{-1/2}$. It is important to
see how the universe becomes scale invariant with the expansion of
the universe. Notice that the temporal evolution of $\beta$, and
hence of the spectral index $n_s$, is due to the existence of
$\lambda$, which has a clear origin in the extra space-like
coordinate $\psi$ [see eqs. (\ref{b18}) and below]. This result
cannot be found in standard 4D inflationary models.

\section{Final Comments}

We have studied a non-perturbative formalism for scalar metric
fluctuations from a 5D extended Schwarzschild-de Sitter (SdS)
cosmological metric. In particular, we have studied the
large-scale cosmological limit, which is the relevant for the
study of large-scale structure formation in the early inflationary
universe. In this limit the intensity of the fluctuations are weak
enough, so we can take a longitudinal gauge, and thus one can
implement the approximation $e^{\pm 2\Phi} \simeq 1\pm 2\Phi$. We
have shown that the energy density on the 4D hypersurface is
induced geometrically by the foliation $\psi=\psi_0=1/H$, in
virtue of the background energy density coming from the inflaton
field being null. This is an important result because all the
contributions of the inflaton field are in the quantum
fluctuations. We have calculated the spectral index for the
scale-invariant metric fluctuations. Our results show how the
universe becomes gradually scale invariant. The time when
inflation ends $\tau_e$ can be calculated from experimental data.
The range of possible values was obtained in (\ref{ta}). Finally,
our 5D model resolves the problem of some 4D standard inflationary
models in which the background values of the inflaton field
$\bar{\varphi}_b$ are transplanckians. In our case
$\bar{\varphi}_b=0$, so that all the background energy density is
due to the 4D induced cosmological constant, which is induced on
the 4D hypersurface by the foliation $\psi=\psi_0=1/H$.

\section*{Acknowledgements}

\noindent J. E. Madriz-Aguilar,  L. M. Reyes and C. Moreno  acknowledge CONACYT (Mexico) and Mathematics Department
of CUCEI- UdG for financial support. M. Bellini acknowledges CONICET (Argentina) and UNMdP for financial support.

\bigskip

\begin{thebibliography}{99}
\bibitem{1} A. H. Guth, Phys. Rev. {\bf D23}: 347(1981).
\bibitem{2} D. H. Lyth and A. Riotto, Phys. Rept. {\bf 314}: 1(1999).
\bibitem{bran} R. H. Brandenberger, Lec. Notes Phys. {\bf 738}, (2008) 393-424.
\bibitem{nrr} Kei-ichi Maeda, Nobuyoshi Ohta, Phys.
Lett. {\bf B597} (2004) 400-407.
\bibitem{we} P. S. Wesson. Gen. Rel. Grav. {\bf 35}: 111(2003).
\bibitem{..} J. E. Madriz Aguilar, M. Bellini, Phys. Lett. {\bf B679}: 306(2009).
\bibitem{...} J. E. Madriz Aguilar, M. Bellini, {\bf JCAP 1011}:
020(2010); \\
L. M. Reyes, J. E. Madriz Aguilar, M. Bellini. Eur. Phys. J. Plus
{\bf 126}:  56(2011).
\bibitem{....} L. M. Reyes, C. Moreno, J. E. Madriz Aguilar, M. Bellini. Phys. Lett. {\bf B717}: 17(2012).
\bibitem{vi} Another approach for a SdS cosmological model of the
    early universe (but developed in a 4D model), can be found in:
    Tomislav Prokopec, Paul Reska. JCAP {\bf 11}: 050(2011).
\bibitem{ga1} J. E. Madriz-Aguilar and M. Bellini, Phys. Lett. {\bf B679}: 306(2009).
\bibitem{pc} T. Shiromizu, D. Ida and T. Torii, JHEP {\bf 11}: 010(2001).
\bibitem{cop} E. J. Copeland, E. W. Kolb, A. R. Liddle and J. E. Lidsey, Phys. Rev. {\bf D 48}: 2529(1993).
\bibitem{rpp} O. Lahav and A. R. Liddle. Phys. Rev. {\bf D80}: 280
(2012).
\end{thebibliography}
\end{document}